\documentclass[12pt]{article}
\usepackage{epsfig,cite}
\def\bea{\begin{eqnarray}}
\def\eea{\end{eqnarray}}
\def\beq{\begin{equation}}
\def\eeq{\end{equation}}
\catcode`@=11 
\def\slash#1{\mathord{\mathpalette\c@ncel#1}}
 \def\c@ncel#1#2{\ooalign{$\hfil#1\mkern1mu/\hfil$\crcr$#1#2$}}
\def\lsim{\mathrel{\mathpalette\@versim<}}
\def\gsim{\mathrel{\mathpalette\@versim>}}
 \def\@versim#1#2{\lower0.2ex\vbox{\baselineskip\z@skip\lineskip\z@skip
       \lineskiplimit\z@\ialign{$\m@th#1\hfil##$\crcr#2\crcr\sim\crcr}}}
\catcode`@=12 

\def\as{\alpha_s}

\relax

\def\abs#1{\left|#1\right|}

\def    \hepph  #1 {{\tt hep-ph/#1}}
\def    \hepex  #1 {{\tt hep-ex/#1}}

\newcommand\rs{\scriptstyle\rm}
\newsavebox\tmpfig

\begin{document}

\pagestyle{empty}

\begin{flushright}

IFUM-860-FT\\ GeF/TH/14-05\\
\end{flushright}

\begin{center}
\vspace*{0.5cm}
{\Large \bf Borel resummation of soft gluon radiation\\ and higher twists}
 \\
\vspace*{1.5cm}
Stefano~Forte,$^{a,b}$ Giovanni~Ridolfi,$^{c}$
Joan~Rojo$^{d}$
and Maria~Ubiali$^{a}$\\
\vspace{0.6cm}  {\it
{}$^a$Dipartimento di Fisica, Universit\`a di Milano,\\
Via Celoria 16, I-20133 Milano, Italy\\ \medskip
{}$^b$INFN, Sezione di Milano,\\
Via Celoria 16, I-20133 Milano, Italy\\ \medskip
{}$^c$Dipartimento di Fisica, Universit\`a di Genova and
INFN, Sezione di Genova,\\
Via Dodecaneso 33, I-16146 Genova, Italy\\ \medskip
{}$^d$Departament d'Estructura i Constituents de la Materia\\
Universitat de Barcelona, Diagonal 647, 08028 Barcelona, Spain}\\
\vspace*{1.5cm}

{\bf Abstract}
\end{center}

\noindent
We show that the well-known divergence of the perturbative expansion
of resummed results for processes such as deep-inelastic scattering
and Drell-Yan in the soft limit can be treated by Borel
resummation. The divergence in the Borel inversion can be removed by
the inclusion of suitable higher twist terms. This provides us with an
alternative to the standard 'minimal prescription' for the 
asymptotic summation of the perturbative expansion, and it gives us some
handle on the role of higher twist corrections in the soft
resummation region.

\vspace*{1cm}

\vfill
\noindent

\begin{flushleft} December 2005 \end{flushleft}
\eject

\setcounter{page}{1} \pagestyle{plain}

It has been known for some time~\cite{bassetto} that the resummation
of logarithmically enhanced contributions to the coefficients of the
QCD perturbative expansion due to soft gluon radiation has the effect
of rescaling the argument of the strong coupling constant: the hard
perturbative scale is replaced by a relatively soft scale related to
the radiation process.
For example, for a physical process
characterized by the hard scale $Q^2$ and a scaling variable $0\le
x\le 1$  near the $x=1$ boundary of phase space (e.g. close to the 
production threshold for a given final state),  the
resummation of large logs of $1-x$ effectively replaces the
perturbative coupling $\as(Q^2)$ with $\as(Q^2(1-x))$.
Similarly, in the resummation of soft $p_T$ spectra~\cite{parisi} the
argument of the strong coupling becomes $p_T^2$, and so on.
Therefore, the perturbative approach eventually fails in this `soft'
kinematical region. This failure is understandable on physical
grounds, because at the phase space boundary, i.e.  as $x\to1$, the
center--of--mass energy is just sufficient to produce the given final
state, so, for instance in the case of deep-inelastic scattering in
this limit the process becomes elastic.

In practice, this problem must be treated in some way in order to
obtain phenomenological predictions: eventually, at some low scale
$\Lambda$ (the position of the Landau pole) the strong coupling blows
up, so when
\beq
x=x_L\equiv 1- \frac{\Lambda^2}{Q^2}\label{xldef}
\eeq
resummed results become meaningless. The scale $\Lambda$ is usually
identified with $\Lambda_{\rm QCD}$.
However, 
$x$--space resummed results are known to run into difficulties anyway,
regardless of the size of the coupling constant,
essentially because the resummation of leading (or
next$^k$--to--leading) logarithmic contributions in $x$ space does not
respect momentum conservation~\cite{cmnt}: this produces a spurious
factorial divergence of resummed results when expanded at fixed
perturbative order. Because of this
factorial divergence, attempts to remove
the problem of the Landau pole by cutting off the resummed $x$--space
result at large $x\lsim x_L$~\cite{cutoff}
display a sizable dependence on the choice of cutoff. It has therefore been suggested~\cite{cmnt} that it is in
fact more advisable to consider resummed results in 
terms of the variable $N$ which is
Mellin conjugate to $x$, namely, to consider the resummation of large
$\ln N$ as
$N\to\infty$. Indeed, it can be shown that to any logarithmic 
order~\cite{fr} upon inverse
Mellin transformation the resummation
of next$^k$-to-leading $\ln N$ contributions provides 
resummation
of next$^k$-to-leading $\ln (1-x)$  terms, up to subleading
contributions. 

If the $N$--space resummed result is expanded
perturbatively in powers of $\alpha_s(Q^2)$ and then Mellin transformed
back to $x$ order by order, one winds up with a divergent series: in
other words, the $N$--space resummed result cannot be obtained as the
Mellin transform
of a perturbative $x$--space calculation. However, it is possible to
give a ``minimal prescription''~\cite{cmnt} for the reconstruction
from the $N$--space resummed result of an
$x$--space result to which this divergent sum is asymptotic
(in the sense of asymptotic series). The minimal prescription (MP)
thus leads to a result which is well--defined and smooth for all $0\le
x\le 1$. It has been widely used for phenomenological applications. 

Here, we wish to reconsider this issue, partly motivated by the
results of ref.~\cite{fr}, which give us full control on the relation
between large $x$ and large $N$ resummations, and by the observation
that the fact that the 
minimal prescription is well defined for all $x$ is in fact a
mixed blessing, if one cannot control what is happening when $x\gsim
x_L$, where the perturbative approach breaks down. We will see that
the divergence of the $x$--space perturbative result can be traced to
subleading terms when  the Mellin inversion of the $N$--space result
is performed to all logarithmic orders (but excluding subleading
powers), and that it can be treated by Borel resummation, at the
expense of including higher--twist contributions. The result which is
obtained shares the pleasing features of the minimal prescription: it
provides an asymptotic summation of the divergent perturbative
expansion, and
it is well--defined for all $x$. However, it differs from it, though
this difference only becomes significant for
$x \sim x_L$ i.e. around the Landau pole (where it is in fact closer
to the truncated perturbative result). Also, it has a rather different
physical interpretation: if suitable higher twist contributions are
included, the perturbative expansion becomes convergent, and our
result provides its sum.

We start by recalling how the Landau pole
appears and is treated with  the minimal prescription. 
Consider a generic observable
$\sigma(Q^2,x)$ and
its Mellin transform
\beq
\label{mellin}
\sigma(Q^2,N)=\int_0^1 dx\,x^{N-1}\,\sigma(Q^2,x),
\eeq
which maps the region of large $x$ onto the region of large $N$.
The resummation of logarithms of $N$ can be performed~\cite{fr} in terms 
of the physical anomalous dimension
\beq
\gamma(\as(Q^2),N)=\frac{\partial\ln\sigma(Q^2,N)}{\partial\ln Q^2}.
\eeq
For example, in the case of deep-inelastic scattering structure functions
the resummed expression of $\gamma(\as(Q^2),N)$ has the
form~\cite{fr,cnt,sterman}
\beq
\gamma(\as(Q^2),N)=
\int_1^N\frac{dn}{n}\,\sum_{k=1}^\infty g_k\,\as^k(Q^2/n)+O(N^0),
\label{gamma}
\eeq
where $g_k$ are constants to be determined by matching with
fixed-order calculations, and the neglected terms are either $N$--independent,
or suppressed by inverse powers of $N$ for large $N$.
Truncating the sum at $k=1,2,\ldots$ corresponds to computing
$\gamma(\as(Q^2),N)$ to leading, next-to-leading, $\ldots$ logarithmic
accuracy.\footnote{Beyond next-to-leading log, eq.~(\ref{gamma}) only
holds provided the cross section has a particular factorization
property. This is immaterial for the ensuing discussion, where we will
concentrate on the leading logarithmic case.} Starting from the resummed expression
of $\gamma(\as(Q^2),N)$, one can obtain the Mellin transform of the cross section,
resummed to the same logarithmic accuracy. The physical quantity
$\sigma(Q^2,x)$ may be obtained by inversion of the Mellin
transform.

Equation~(\ref{gamma}) shows explicitly that the resummed result depends
on $\as(Q^2/N)$: the resummation has replaced the hard scale $Q^2$
with the softer scale $Q^2/N$ --- in fact, in the soft limit 
the resummed result only
depends on $N$ through this rescaled coupling.
As a consequence, $\gamma(\as(Q^2),N)$ has a branch cut along the real positive axis
for $N>N_L$, where $N_L$ is the location of the $N$--space Landau pole
\beq
\label{nlandau}
N_L=\frac{Q^2}{\Lambda^2}.
\eeq
This in particular implies that the inverse Mellin transform of
the resummed anomalous dimension eq.~(\ref{gamma}) does not exist. Indeed,
the Mellin transform of a function $f(x)$ such that
$\abs{f(x)}<Kx^{- N_0}$ for all $x$, with $K$ and
$N_0$ real constants, is an analytic function
of the complex variable $N$ in the half-plane ${\rm Re\,}N>N_0$. Therefore, $\gamma(\as(Q^2),N)$  eq.~(\ref{gamma})  cannot be the Mellin
transform of any function. 

To see the problem more clearly,  let us consider for definiteness
the resummed expression of
$\gamma(\as(Q^2),N)$ to leading logarithmic accuracy,
\beq 
\gamma_{\rs LL}(\as(Q^2),N)=g_1 \int_1^N\frac{dn}{n}\,\as(Q^2/n)
=-\frac{g_1}{\beta_0}
\ln\left(1+\beta_0\as(Q^2)\ln\frac{1}{N}\right),
\label{gammaLL}
\eeq
where we have consistently used the leading-log expression of $\as$:
\beq
\as(\mu^2)=\frac{\as(Q^2)}{1+\beta_0\as(Q^2)\ln\frac{\mu^2}{Q^2}};
\qquad \beta_0=\frac{33-2n_f}{12\pi}.
\eeq
Because of the Landau singularity,
$\gamma_{\rs LL}(\as(Q^2),N)$ has a branch cut on the real
positive axis for 
\beq
\label{nbranch}
N\geq N_L\equiv e^\frac{1}{\beta_0\as(Q^2)}.
\eeq

One may formally consider the term-by-term inverse Mellin transform of
the expansion of $\gamma_{\rs LL}(\as(Q^2),N)$ in powers of $\as(Q^2)$.
This gives
\beq
\label{Pasympt}
P_{\rs LL}(\as(Q^2),x)=-\lim_{K\to\infty}\frac{g_1}{\beta_0}
\sum_{k=1}^K \frac{(-1)^{k+1}}{k}\beta_0^k \as^k(Q^2)\,
\frac{1}{2\pi i}\int_{\bar N-i\infty}^{\bar N+i\infty}dN\,
x^{-N}\,\ln^k\frac{1}{N};\qquad \bar N>0.
\eeq
Each term of the series is now a well-defined inverse Mellin transform,
but the series does not converge, so we cannot take the limit $K\to\infty$.  Indeed, if the series were convergent,
one could interchange the sum over $k$ and the integral over $N$ in eq.~(\ref{Pasympt}), but the sum
\beq
\sum_{k=1}^\infty \frac{(-1)^{k+1}}{k}\beta_0^k \as^k(Q^2)\,
\ln^k\frac{1}{N}
\eeq
is only convergent for
\beq
\abs{\beta_0\as(Q^2)\ln\frac{1}{N}}<1,
\eeq
while the integral in $N$ on the path ${\rm Re\,}N=\bar N$
involves values of $N$ outside this range.

However, as well known, if the Mellin transform in eq.~(\ref{Pasympt}) is
computed at the relevant (leading, next-to-leading...) logarithmic level
the perturbative series converges. Indeed,
considering again for definiteness the leading log
case one has 
\beq
\frac{1}{2\pi i}\int_{\bar N-i\infty}^{\bar N+i\infty}dN\,
x^{-N}\,\ln^k\frac{1}{N}
=k\,\left[\frac{\ln^{k-1}(1-x)}{1-x}\right]_+ +{\rm NLL},
\label{IL}
\eeq
where $+$ denotes the standard prescription of Altarelli-Parisi
evolution~\cite{ap}.
The series in eq.~(\ref{Pasympt}) now is convergent for all $x<x_L$:
its sum is
\beq
\label{PLL}
P_{\rs LLx}(\as(Q^2),x)=
-g_1\left[\frac{1}{1-x}\frac{\as(Q^2)}{1+\beta_0\as(Q^2)\ln(1-x)}\right]_+
=-g_1\left[\frac{\as(Q^2(1-x))}{1-x}\right]_+,
\eeq
which is singular at the Landau pole eq.~(\ref{xldef}). This
singularity sets the radius of convergence of the series.  Similar
arguments can be used to show~\cite{fr} that if we start with the
next$^k$-to-leading $\ln\frac{1}{N}$ result and we perform the Mellin
inversion to the same order we wind up with a series of terms of the
form of eq.~(\ref{PLL}), but with higher powers of the coupling, up to
$\as^k(Q^2(1-x))$.  Equation~(\ref{PLL}) shows explicitly that the
resummation replaces the scale $Q^2$ with $Q^2(1-x)$. Notice that
eq.~(\ref{PLL}) and its higher--order cognates provide us with a
next$^k$-to-leading $\ln(1-x)$ resummation at the level of the
physical anomalous dimension (or rather splitting function), rather
than of the cross section, and it is free of the spurious factorial
growth mentioned above. In fact, it can be shown that the factorial
growth is a byproduct of the next$^k$-to-leading log truncation of the
exponentiated result, and it disappears provided only the Mellin
transform of the exponentiated result is determined including
subleading logarithmic corrections to all orders~\cite{fr}: it is
therefore totally unrelated~\cite{cmnt} to the problem of the Landau
singularity discussed here, and we will not worry about
it further.

The minimal prescription, proposed in ref.~\cite{cmnt}, 
consists of {\it defining} $P_{\rs LL}(\as(Q^2),x)$ as an integral along a
contour that passes to the left of the Landau pole:
\beq
\label{PMP}
P_{\rs LL}^{\rs MP}(\as(Q^2),x)=
\frac{1}{2\pi i}\int_{N_{\rs MP}-i\infty}^{N_{\rs MP}+i\infty}dN\,x^{-N}\,
\gamma_{\rs LL}(\as(Q^2),N);\qquad 0<N_{\rs MP}<N_L.
\eeq
Notice that, because of the  branch cut eq.~(\ref{nbranch}), integration
to the right of the Landau pole is in fact not possible. 
The
function $P_{\rs LL}^{\rs MP}(\as(Q^2),x)$ is then free of Landau
singularities. Furthermore, as proved in ref.~\cite{cmnt} 
the series in
eq.~(\ref{Pasympt}), despite being divergent, is asymptotic
to $P_{\rs LL}^{\rs MP}(\as(Q^2),x)$: the difference between the minimal
prescription result eq.~(\ref{PMP}) and the $k$-th order truncation
of the divergent series eq.~(\ref{Pasympt}) is
$O(\as^{k+1})$. Interestingly, the remainder grows less than
factorially (as $(\ln k)^k$ for large $k$).

We would like instead to tackle directly the divergent perturbative series
eq.~(\ref{Pasympt}).
The Mellin inversion integral can be computed explicitly:
\bea
\frac{1}{2\pi i}\int_{\bar N-i\infty}^{\bar N+i\infty}dN\,
x^{-N}\,L^k
&=&\left.\frac{d^k}{d\eta^k}\,
\frac{1}{2\pi i}\int_{\bar N-i\infty}^{\bar N+i\infty}dN\,
x^{-N}\,N^{-\eta}\right|_{\eta=0}\nonumber\\
&=&\frac{d^k}{d\eta^k}\,
\frac{1}{\Gamma(\eta)}\left[\ln^{\eta-1}\frac{1}{x}\right]_+
\Bigg|_{\eta=0}\nonumber +\delta(1-x)\,
\label{Bconv1}
\eea
where the last equality follows from the identity $\int_0^1dx\,x^{N-1}\,
\left[ \ln^{\eta-1}\frac{1}{x}\right]_+=\Gamma(\eta)(N^{-\eta}-1).$
Hence,
\bea
P_{\rs LL}(\as(Q^2),x)&=&\frac{g_1}{\beta_0}
\sum_{k=0}^K \frac{[-\beta_0 \as(Q^2)]^{k+1}}{k+1}\,
\Bigg\{\sum_{n=0}^{k+1}
\left(\begin{array}{c}k+1\\n\end{array}\right)
\left(\frac{d^n}{d\eta^n}\,\frac{1}{\Gamma(\eta)}\right)
\frac{d^{k+1-n}}{d\eta^{k+1-n}}
\left[\ln^{\eta-1}\frac{1}{x}\right]_+\Bigg|_{\eta=0}
\nonumber\\
&&
\qquad+\delta(1-x)\Bigg\}
\nonumber\\
&=&\frac{g_1}{\beta_0}
\sum_{k=0}^K \frac{[-\beta_0 \as(Q^2)]^{k+1}}{k+1}
\Bigg\{\left[\frac{1}{\ln\frac{1}{x}}\sum_{n=1}^{k+1}
\left(\begin{array}{c}k+1\\n\end{array}\right)
 n \Delta^{(n-1)}(1) \left( \ln\ln\frac{1}{x}\right)^{k+1-n}\right]_+
\nonumber\\
&&\qquad+\delta(1-x)\Bigg\},
\label{Pasymptexp}
\eea
where in the last step we have defined $\Delta(z)\equiv1/\Gamma(z)$, and we
have used the identity
$\Delta^{(k)}(0)=k\Delta^{(k-1)}(1)$.
%
With straightforward manipulations we get
\bea
P_{\rs LL}(\as(Q^2),x)&=&\frac{g_1}{\beta_0}
\sum_{n=0}^K[-\beta_0 \as(Q^2)]^{n+1}\,
\Bigg\{
\frac{\Delta^{(n)}(1)}{n!}
\left[\frac{1}{\ln\frac{1}{x}}\sum_{k=n}^K
\frac{k!}{(k-n)!}\,
[-\beta_0 \as(Q^2)\ln\ln\frac{1}{x}]^{k-n}\right]_+
\nonumber\\
&&
+\frac{1}{n+1}\,\delta(1-x)\Bigg\}
\nonumber\\
&=&
\frac{g_1}{\beta_0}
\sum_{n=0}^K\Bigg\{\Delta^{(n)}(1)\,
\left[\frac{1}{\ln\frac{1}{x}}
[-\beta_0\as(Q^2\ln\frac{1}{x})]^{n+1}\right]_+
+\frac{[-\beta_0 \as(Q^2)]^{n+1}}{n+1}\,\delta(1-x)\Bigg\}
\nonumber\\
&&+O(\as^{K+1}).
\label{Pasymptsum}
\eea
In the limit $K\to\infty$ the terms of order $\as^{K+1}$ can be
neglected, but the series is divergent.

In the large $x$ limit, $\ln\frac{1}{x}=1-x +O((1-x)^2)$, so, to logarithmic
accuracy we may rewrite
eq.~(\ref{Pasymptsum}) as 
\beq
P_{\rs LL}(\as(Q^2),x)=\frac{g_1}{\beta_0}
\left[\frac{R(\as(Q^2),x)}{1-x}\right]_+
\label{lxdiv}
\eeq
where
\beq
R(\as(Q^2),x)=\lim_{K\to\infty}
\sum_{n=0}^K\Delta^{(n)}(1)\,[-\beta_0 \as(Q^2(1-x))]^{n+1},
\label{PasymptLL}
\eeq
which holds up to non-logarithmic terms. Note that
eq.~(\ref{PasymptLL}) only follows from eq.~(\ref{Pasymptsum}) when
$K\to\infty$: even when $K$ is finite an infinite number of terms in
eq.~(\ref{Pasymptsum}) is needed in order to reconstruct 
$\as(Q^2(1-x))$. 
Equation~(\ref{PasymptLL}) can also be obtained directly by
computing the inverse Mellin transform eq.~(\ref{Bconv1}) to
logarithmic accuracy, thanks to the result of ref.~\cite{fr}
\beq
\frac{1}{2\pi i}\int_{\bar N-i\infty}^{\bar N+i\infty}dN\,
x^{-N}\,L^{k+1}
=\sum_{n=0}^k\left(
\begin{array}{c}k+1\\n\end{array}\right)
\Delta^{(n)}(1)\,(k+1-n)\left[\frac{\ln^{k-n}(1-x)}{1-x}\right]_+.
\label{Lk}
\eeq
Equation~(\ref{PasymptLL}) shows that the divergence of the series
eqs.~(\ref{Pasympt},\ref{Pasymptsum}) is due to the Mellin inversion of
$\ln N$ to all logarithmic orders: if  eq.~(\ref{Lk})
is truncated to any finite logarithmic order, the resummed result in $x$ space
converges, with finite radius $x<x_L$, as in the
leading $\ln(1-x)$ case, eq.~(\ref{PLL}), but if all logarithmic orders
are included, then the series diverges, and the inclusion of power
suppressed terms does not bring in any new divergence. 

Having understood the origin of the divergence, we can now proceed to
its summation by the Borel method. Since we are interested in the
large $x$ limit, we neglect power-suppressed terms, and we use the
all-log result eq.~(\ref{PasymptLL}).  Namely, we take the Borel
transform of the divergent series (\ref{PasymptLL}) with respect to
$\beta_0 \as(Q^2(1-x))$, thereby obtaining the Taylor series expansion
of the function $\Delta(z)$ about $z=1$:
\beq
\hat R(w,x)=-\sum_{j=0}^\infty\frac{\Delta^{(j)}(1)}{j!}
\,(-w)^j=-\frac{1}{\Gamma(1-w)}.
\label{PwxLL}
\eeq
Because $\Delta(z)$ is an entire function, the radius of convergence
of the Borel transformed series eq.~(\ref{PwxLL}) is infinite.

The Borel sum of the original series is given by the inverse Borel transform 
\beq
R_{\rs B}(\as(Q^2),x)=-\int_0^{+\infty}dw\,e^{-\frac{w}{\beta_0\as(Q^2(1-x))}}\,
\frac{1}{\Gamma(1-w)}.
\label{invB}
\eeq
However, the integrand  diverges as $w\to\infty$, because the
reflection formula
\beq
\frac{1}{\Gamma(1-w)}=\frac{1}{\pi} \Gamma(w)\sin(\pi w)
\label{ref}
\eeq
implies that $\Delta(1-w)$ oscillates with a factorially growing
amplitude as $w\to\infty$ on the real axis: hence,
the Borel sum is ill-defined. 

One may think that, alternatively, we could
have performed a Borel transform with respect to $-\beta_0
\as(Q^2(1-x))$. In this case, we end up with
\beq
R^-_{\rs B}(\as(Q^2),x)=-\int_0^{+\infty}dw\,e^{\frac{w}{\beta_0\as(Q^2(1-x))}}\,
\frac{1}{\Gamma(1+w)}.
\label{invBm}
\eeq
The integral now converges, because of the factorial damping provided
by $\frac{1}{\Gamma(1+w)}$ as $w\to\infty$.  However,
eq.~(\ref{invBm}) diverges in the limit $\as(Q^2)\to0$ in the physical
region where $\as >0$.  It is amusing to note that in the unphysical
region $\as <0$ it can be easily proved that this Borel summation
coincides with the minimal prescription. However, the physical region
and the unphysical region of the minimal prescription manifestly
cannot be analytically continued into each other, because in the
unphysical region the cut is actually to the left of the path of
integration in eq.~(\ref{PMP}). In the physical region, instead, this
modified Borel result eq.~(\ref{invBm}) is physically unacceptable
because it blows up in the perturbative limit --- and in fact it is
very large even for moderate values of $\as(Q^2)$, because the factor
$e^{\frac{w}{\beta_0\as(Q^2(1-x))}}$ is huge before the damping due to
the factor of $\Gamma(1+w)^{-1}$ sets in. Hence, we conclude that the
result eq.~(\ref{invBm}) is unphysical, and we must stick with the
result eq.~(\ref{invB}).

The presence of singularities along the path of
integration in Borel inversion is a common occurrence in perturbative
QCD, e.g. in the context of renormalons~\cite{renorm}, and it is dealt
with by cutting off the singularity. In our case, the singularity is
as $w\to\infty$, hence we must introduce an upper cutoff $C$ to the
integral. We therefore replace the divergent result eq.~(\ref{invB}) by
\beq
R_{\rs B}(\as(Q^2),x,C)=
-\int_0^C d w\,
\frac{1}{\Gamma(1-w)}\,
e^{-\frac{w}{\beta_0\as(Q^2(1-x))}},
\label{pbreg}
\eeq
which is  convergent for all finite $C$. The regulated result
eq.~(\ref{pbreg}) is well defined for all $x$. Indeed,
if we expand the integrand according to
eq.~(\ref{PwxLL}), the
series converges uniformly over the
integration range for all finite $C$, so we may integrate 
term by term, with the result
\beq
R_{\rs B}(\as(Q^2),x,C)=
\sum_{k=0}^\infty \Delta^{(k)}(1)
[-\beta_0\as(Q^2(1-x))]^{k+1} f_k
\label{series}
\eeq
where
\beq
f_k\equiv
\frac{\gamma\left(k+1,\frac{C}{\beta_0\as(Q^2(1-x))}\right)}
{\Gamma(k+1)}
\label{ffact}
\eeq
and
$\gamma(k,z)$ is the truncated Gamma function
\beq
\gamma(k+1,z)\equiv\int_0^z\! dw\, e^{-w} w^k .
\label{truncgamma}
\eeq
The series eq.~(\ref{series}) is convergent; however, if the cutoff $C$
is taken to infinity, $\lim_{C\to\infty}f_k=1$, and 
the original divergent series is reproduced.

It is easy to see  that the sum of 
the convergent series eq.~(\ref{series}) is an
asymptotic sum of the divergent series eq.~(\ref{invB}). Indeed, 
rewrite the series eq.~(\ref{series}) as 
\beq
R_{\rs B}(\as(Q^2),x,C)=R_{lt}(\as(Q^2),x)-R_{ht}(\as(Q^2),x,C),
\label{twdec}
\eeq
where
\beq
R_{lt}(\as(Q^2),x)\equiv\sum_{k=0}^\infty \Delta^{(k)}(1)
[-\beta_0\as(Q^2(1-x))]^{k+1} =R(\as(Q^2),x)
\label{lt}
\eeq
and
\beq
R_{ht}(\as(Q^2),x,C)\equiv e^{-\frac{C}{\beta_0\as(Q^2(1-x))}}
\sum_{k=0}^\infty \Delta^{(k)}(1)
[-\beta_0\as(Q^2(1-x))]^{k+1}   \sum_{n=0}^k \frac{1}{n!}
\left(\frac{C}{\beta_0\as(Q^2(1-x))}  \right)^n.
\label{ht}
\eeq
The
difference between the convergent series eq.~(\ref{series}) and the
first $k_0$ orders of the divergent series eq.~(\ref{invB}) is equal to
the sum of two terms. The first is $R_{lt}$ with only terms with
$k>k_0$ included. This is of order $\as^{k_0+1}$. The second is the
$R_{ht}$, which is proportional to
$\exp\left[-1/(\beta_0\as)\right]$, 
and therefore as $\as\to0$  it vanishes faster
than any power of $\as$. The remainder of the
asymptotic sum grows like the coefficients $\Delta^{(k)}(1)$ of
eq.~(\ref{lt}). These grow less than factorially (like the remainder
of the minimal prescription), because the Taylor expansion of the
function $\Delta(z)$ has infinite convergence radius. 

The sum $R_{\rs B}(\as(Q^2),x,C)$ of the series eq.~\ref{series} is regular
for all $0\le x \le1$.
In particular,  it is regular at the Landau pole, where it
takes a finite value which depends on $C$. 
 Using the standard leading--order expression of $\alpha_s$
\beq
\as(Q^2)=\frac{1}{\beta_0\ln\frac{Q^2}{\Lambda^2}}
\label{loas}
\eeq
and the identity
\beq
\gamma(k+1,z)=k!\left(1-e^{-z}\sum_{n=0}^k \frac{1}{n!} z^n\right),
\label{truncgammaint}
\eeq
the $k$--th order contribution to the series eq.~(\ref{series})  is
\beq
[-\beta_0\as(Q^2(1-x))]^{k+1} f_k=
\left(-\frac{1}{\ln\frac{Q^2(1-x)}{\Lambda^2}}\right)^{k+1}
e^{-C \ln\frac{Q^2(1-x)}{\Lambda^2}}
\sum_{n=k+1}^\infty \frac{1}{n!} 
\left(C \ln\frac{Q^2(1-x)}{\Lambda^2}\right)^n.
\eeq
In the limit $x\to x_L$, $\ln\frac{Q^2(1-x)}{\Lambda^2}\to 0$, and we get
\beq
[-\beta_0\as(Q^2(1-x_L))]^{k+1} f_k=
\frac{(-1)^{k+1}}{(k+1)!} C^{k+1}.
\eeq
Hence
\beq
R_{\rs B}(\as(Q^2),x_L,C)=\sum_{k=0}^\infty \Delta^{(k)}(1)\,
\frac{(-1)^{k+1}}{(k+1)!} C^{k+1}=-\int_0^C d\eta\,\Delta(1-\eta).
\eeq
The same result is obtained by simply taking the limit $x\to x_L$
in eq.~(\ref{pbreg}).

\begin{figure}[t]
\begin{center}
\epsfig{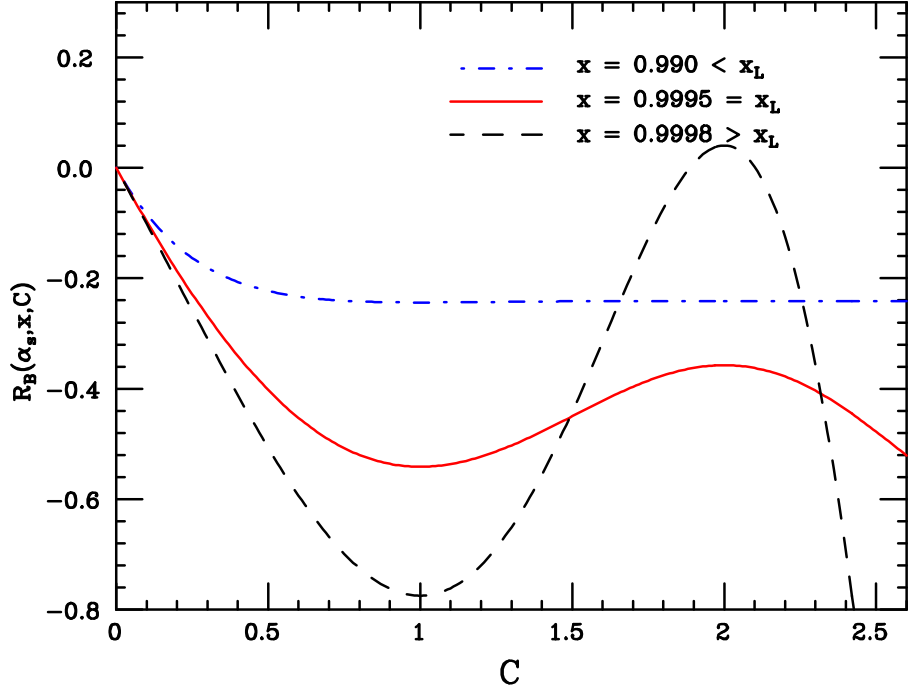}
\end{center}
\begin{center}
\caption{\small Dependence of the truncated Borel integral $R_{\rs B}$
eq.~(\ref{pbreg}) on the cutoff $C$ for $\as(Q^2)=0.25$ and 
three values of $x$: below, at and above the Landau pole.}
\label{fig:cutdep}
\end{center}
\end{figure}
\begin{figure}[t]
\begin{center}
\epsfig{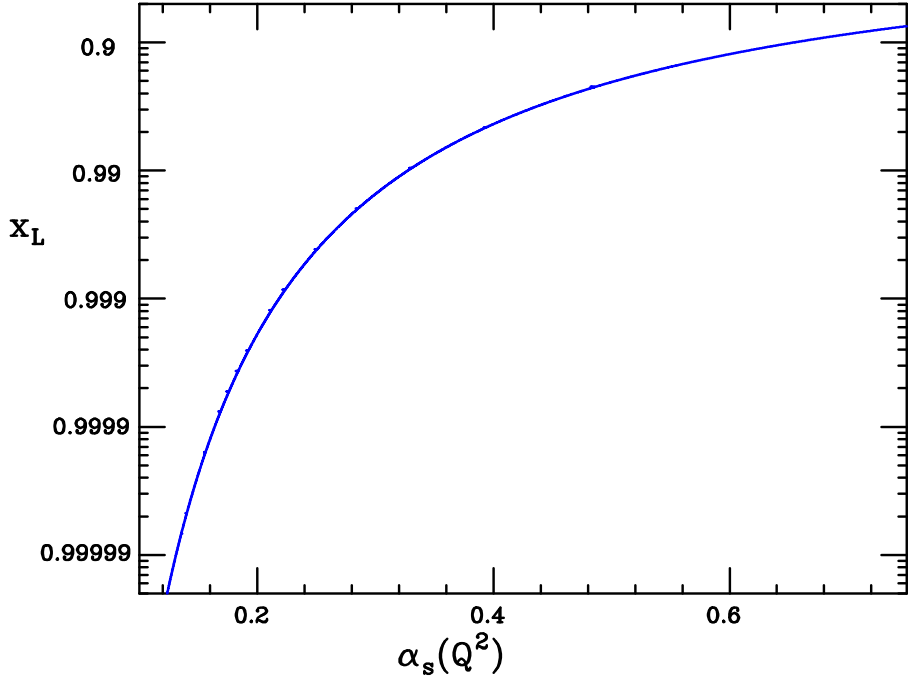}
\end{center}
\begin{center}
\caption{\small Position of the leading--order Landau pole
eq.~(\ref{polepos}) as a function of $\alpha_s(Q^2)$.}
\label{fig:polepos}
\end{center}
\end{figure}

The dependence of the value of $R_{\rs B}(\as(Q^2),x,C)$
on $C$ is an $O(\exp-\frac{1}{\as})$ ambiguity in the definition of
the asymptotic sum of the divergent series. 
This dependence is displayed in  fig.~\ref{fig:cutdep} for three
values of $x$, below, at, and above the Landau pole, which 
at leading order is located at
\beq
x_L=1- e^{-\frac{1}{\beta_0\as(Q^2)}}.
\label{polepos}
\eeq
For ease of reference, the dependence of the position $x_L$ of the
Landau pole on the value of $\as(Q^2)$ is 
displayed in fig.~\ref{fig:polepos}.
Firstly, it is clear from eq.~(\ref{pbreg}) that, because 
\beq
\frac{\partial}{\partial C} R_{\rs B}(\as(Q^2),x,C)=
-\frac{1}{\Gamma(1-C)}\,
e^{-\frac{C}{\beta_0\as(Q^2(1-x))}},
\label{cutder}
\eeq
the integral has its first stationary point at $C=1$, and then it is
stationary at all positive integer values of $C$, where $\Gamma(1-C)$
has simple poles. The first two stationary points are clearly visible
in the figure.  For $x<x_L$, however, $R_{\rs B}(\as(Q^2),x,C)$ has a
plateau for $C\gsim 1$. The origin of this plateau is clear from
inspection of eq.~(\ref{cutder}) again: using eq.~(\ref{ref}) it is
apparent that $\left|\frac{1}{\Gamma(1-C)}\right|\lsim 1$ for $C\gsim
1$, whereas as soon as $x<x_L$, $\beta_0\as(Q^2(1-x))<1$ in the
perturbative (large $Q^2$) region, so
$e^{-\frac{C}{\beta_0\as(Q^2(1-x))}}\ll 1$. It is only when
$C\sim\exp\frac{1}{\beta_0\as(Q^2(1-x))}\gg1$ that the factorial
growth catches up with the exponential damping. Up to this value of
$C$ the growth of $R_{\rs B}(\as(Q^2),x,C)$ with $C$ is negligible.
This is as it should be: when $x$ is below the Landau pole,
the asymptotic sum of the series is
essentially independent of the value of $C$, unless one chooses an
unnaturally large or small value. At the Landau pole, a dependence on
the value of $C$ appears, due to the fact that if $x\gsim x_L$ 
the exponential $O(\exp(1/\as))$  
prefactor $e^{-\frac{C}{\beta_0\as(Q^2(1-x))}}\gsim 1$,
and the concept of
asymptotic sum starts losing its meaning. A minimal choice of $C$ 
is $C=1$, so $R_{\rs B}$ is stationary for all $x$, and in
the beginning of the plateau for $x<x_L$. 

The physical meaning of $C$ becomes clear by rewriting
\beq
e^{-\frac{C}{\beta_0\as(Q^2(1-x))}}=
\left(\frac{\Lambda^2}{Q^2(1-x)}\right)^C.
\label{htdef}
\eeq
In other words, $R_{ht}(\as(Q^2),x,C)$ eq.~(\ref{ht}) is a higher twist
contribution: 
the original divergent perturbative series
eq.~(\ref{lt}) has been made convergent by the inclusion of the
higher twist series eq.~(\ref{ht}). Of course, eq.~(\ref{twdec})
implies that this
twist series must necessarily also be divergent. 
Integer values of $C$ correspond to
even twists, and in particular if $C=1$,  $R_{ht}(\as(Q^2),x,C)$ 
is a standard twist--4
contribution, namely, the first subleading twist. 
The
choice $C=1$ is minimal in that it corresponds to regulating the Borel
summation through the first subleading twist. 

Equation~(\ref{htdef}) implies that  these higher
twist terms are suppressed by powers of $\frac{\Lambda^2}{Q^2}$, but
enhanced by powers of $\frac{1}{1-x}$. The Landau pole is the point
where the parameter of the twist expansion is equal to one, i.e.,
leading twist and higher twist terms are of comparable size.
However, despite this enhancement, 
as long as  we choose $C\leq1$  the Borel sum eq.~(\ref{pbreg}) remains
integrable at $x=1$. Indeed, with $C=1$ we have
\beq
(1-x) R_{\rs B}(\as(Q^2),x,1)=
-\int_0^1 d w\,
\frac{1}{\Gamma(1-w)}\,
e^{-\frac{w}{\beta_0\as(Q^2)}} (1-x)^{1-w},
\label{pbint}
\eeq
which vanishes as $x\to 1$ thanks to the fact that $0\le w\le 1$. It
follows that $P_{\rs LL}$ eq.~(\ref{lxdiv}) with $R$ given by $R_{\rs B}$
eq.(\ref{pbreg}) acts as a conventional $+$ distribution, and in
particular the integral 
 $\int_0^1 P_{\rs LL}(\as(Q^2),x) f(x)dx$ is finite if
the test function $f(x)$ is regular at $x=1$. 

When
$R_{ht}$ is viewed as a genuine higher twist term, the prefactor
$\frac{\Lambda^2}{Q^2}$ in eq.~(\ref{ht}) comes from the Wilson
expansion, and not from a factor of $\exp\frac{1}{\as}$, so the higher
twist term is just of $O(\as)$.  However, this term matches an
ambiguity in the leading twist, which does not appear at any finite
order in the expansion of the leading twist term itself but only in
its asymptotic Borel resummation.
Equivalently, the higher twist contribution removes the cutoff
ambiguity introduced by the need to treat this divergence. The
situation is thus akin to the customary case of renormalons, where
similarly the ambiguity introduced by the need to make the Borel
inversion well--defined is cured by the inclusion of higher twist
terms. 
Henceforth, we will take our result to be given by
the Borel sum eq.~(\ref{pbreg}) with $C=1$.  

\begin{figure}[t]
\begin{center}
\epsfig{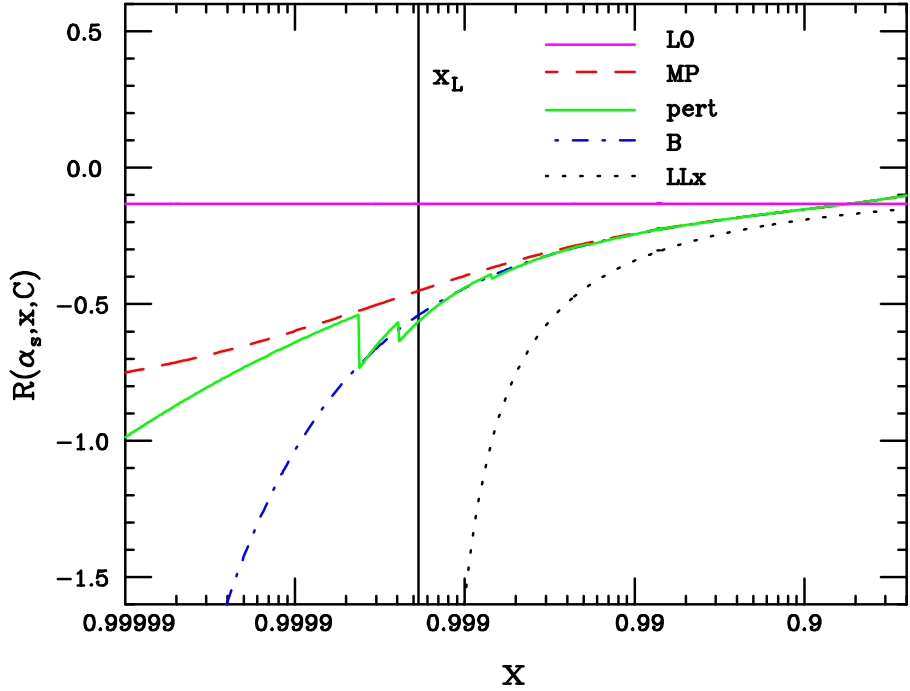}
\end{center}
\begin{center}
\caption{\small Various determination of the $x$--space resummed
result for $\as(Q^2)=0.25$. LLx denotes the leading $\ln(1-x)$ result of
eq.~(\ref{PLL}), while B, MP and pert denote three different
determinations of the divergent leading $\ln N$ series
eq.~(\ref{PasymptLL}), respectively through Borel summation
eq.~(\ref{pbreg}) with $C=1$, the minimal prescription eq.~(\ref{PMP})
and the asymptotic truncation of the perturbative expansion at $K_0$
eq.~(\ref{astrunc}). The large--$x$ (constant) leading order result
eq.~(\ref{lo}) is also shown for comparison.  }
\label{fig:comp}
\end{center}
\end{figure}
We now wish to compare this result to the minimal prescription. We do
so by defining 
\beq
R_{\rs MP}(\as(Q^2),x)\equiv (1-x) P_{\rs MP}(\as(Q^2),x),
\label{rmpdef}
\eeq
where $P_{\rs MP}(\as(Q^2),x)$ is given by eq.~(\ref{PMP}).
It is clear that the 
Borel asymptotic sum eq.~(\ref{pbreg}) and the 
minimal prescription asymptotic sum eq.~(\ref{rmpdef}) of the
divergent series eq.~(\ref{invB})
cannot coincide in general, because the former depends on $C$ and the
latter doesn't. Their $x$--dependence when $C=1$ is compared in
fig.~\ref{fig:comp}, where we also show the value of $R$ obtained 
by truncating the perturbative expansion of the divergent 
series eq.~(\ref{invB}).
This is defined by including in $R$ the contribution from the
first $K_0$ terms in the expansion in powers of $\as(Q^2)$
eq.~(\ref{Pasympt}), where $K_0$ is defined as the value where the  
$k^{th}$ term in the sum in eq.~(\ref{Pasympt}), $a_k$, starts
growing, i.e. as the value $K_0$ such that 
\beq
\abs{a_{k}+a_{k-1}}>\abs{a_{k-2}+a_{k-3}}\quad\hbox{for}\>\> k>K_0.
\label{astrunc}
\eeq
The definition of the optimal truncation point has some degree
of arbitrariness; we have checked that with our choice, eq.~(\ref{astrunc}),
the truncated sum is closer to the asymptotic sum than it would be 
by simply requiring that $\abs{a_{k}}>\abs{a_{k-1}}$ for $k>K_0$. Note
that we have defined the truncation in terms of the expansion
eq.~(\ref{Pasympt}) in powers of $\as(Q^2)$, and not in terms of that
eq.~(\ref{invB}) in powers of $\as(Q^2(1-x))$, in order for the
truncated result to be well-defined also at the Landau pole $x_L$ where
$\as(Q^2(1-x))$ blows up.
For comparison, we also include in fig.~\ref{fig:comp} 
the leading $\ln(1-x)$ result  eq.~(\ref{PLL}), 
and the large $x$ form of the unresummed
result, which is simply given by
\beq
R_{\rs LO}(\as(Q^2),x)= -\beta_0\as(Q^2)+O[(1-x)].
\label{lo}
\eeq

Below the Landau pole the Borel and minimal  resummation prescriptions are
close to the truncated perturbative result and thus close to each
other, as  one expects of an asymptotic sum.  Note that the leading 
 $\ln(1-x)$ result eq.~(\ref{PLL}) is reasonably close to these
results but never quite on top of them: at not so large $x$, where it
reduces to the leading-order result eq.~(\ref{lo}), it differs from
them because of subleading terms --- this is the region were the
large $x$ resummation is not very useful. As one enters the large $x$
region, however, the leading 
 $\ln(1-x)$ result eq.~(\ref{PLL}) is contaminated by the Landau
pole where it blows up. So the leading 
 $\ln(1-x)$ result turns out to be of limited usefulness,
because there is no region of $x$ where it is applicable. 
At and
above the Landau pole the Borel prescription and the MP prescription
start deviating:  in this region the higher twist
contributions which stabilize the Borel sum are of the same order as
the leading twist. On the other hand, at and above the Landau pole the series
diverges very fast and its asymptotic sum looses meaningfulness.
Hence, comparison of the two prescriptions (Borel and minimal) gives
us an estimate of the size of nonperturbative effects: when the two
prescriptions start departing from each other, nonperturbative effects become
important. Indeed, these two prescriptions bracket the truncated
perturbative expansion, which oscillates between them as the order of
the truncation (the  value of $K_0$) varies as a function of $x$.

In summary, we have traced the origin of the divergence of the
perturbative expansion of soft gluon resummation, and we have shown
that it may be treated by Borel resummation stabilized by higher twist
terms. The result that we found is close to the widely adopted minimal
prescription, but it deviates from it when nonperturbative corrections
become important, namely at the Landau pole. All our computations were
presented in the case of threshold resummation (such as e.g. DIS at
large $x$) at the leading logarithmic level. The extension to all
logarithmic orders and to $p_T$ resummation will be discussed elsewhere.
Our result is useful for practical calculations in that it does not
require the numerical evaluation of a Mellin inversion integral. Furthermore,
the availability of more resummation methods that differ in the nonperturbative
region is useful in order to assess the reliability of perturbative
resummed results.
\bigskip

\section*{Acknowledgement}
GR would like to thank Ernesto De Vito for interesting discussions
on the theory of divergent series.

\end{document}